\documentclass[fleqn,usenatbib]{mnras}

\usepackage{newtxtext,newtxmath}

\usepackage[T1]{fontenc}
\usepackage{ae,aecompl}

\usepackage{graphicx}	
\usepackage{amsmath}	
\usepackage{amssymb}	

\usepackage[percent]{overpic}

\usepackage[frozencache,cachedir=.]{minted} 
\usepackage[skins,minted]{tcolorbox}

\definecolor{mintedbackground}{rgb}{0.95,0.95,0.95}

\setminted[python]{
    bgcolor=mintedbackground,
    fontfamily=tt,
    linenos=false,
    numberblanklines=true,
    numbersep=12pt,
    numbersep=5pt,
    gobble=0,
    funcnamehighlighting=true,
    tabsize=4,
    obeytabs=false,
    mathescape=false,
    samepage=false,
    showspaces=false,
    showtabs =false,
    texcl=false,
    baselinestretch=1.1, 
    breaklines=true,
}

\newtcblisting{myminted}[2][]{listing engine=minted, listing only,#1, title=#2, minted language=python, colback=mintedbackground}
\newtcblisting{myminted2}[1][]{listing engine=minted, listing only,#1, minted language=python, colback=mintedbackground}

\title[A seismic scaling relation for stellar age]{A seismic scaling relation for stellar age}

\author[E.\ P.\ Bellinger]{
Earl Patrick Bellinger$^{1}$\thanks{E-mail: bellinger@phys.au.dk}
\\
$^{1}$SAC Postdoctoral Research Fellow, Stellar Astrophysics Centre, Department of Physics and Astronomy, Aarhus University, Denmark
}

\date{Accepted XXX. Received YYY; in original form ZZZ}

\pubyear{2019}

\newif\ifref
\reffalse
\definecolor{darkred}{rgb}{0.7, 0, 0}
\newcommand{\mb}[1]{\ifref\textcolor{darkred}{#1}\else #1\fi}

\newif\ifreff
\refffalse
\definecolor{darkred}{rgb}{0.7, 0, 0}
\newcommand{\mbb}[1]{\ifreff\textcolor{darkred}{#1}\else #1\fi}

\begin{document}
\label{firstpage}
\pagerange{\pageref{firstpage}--\pageref{lastpage}}
\maketitle

\begin{abstract} 
    A simple solar scaling relation for estimating the ages of \mb{main-sequence} stars from asteroseismic and spectroscopic data is developed. 
    New seismic scaling relations for estimating mass and radius are presented as well, including a purely seismic radius scaling relation (i.e., no dependence on temperature). 
    The relations show substantial improvement over the classical scaling relations and perform similarly well to grid-based modeling. 
\end{abstract} 



\section{Introduction} 
A solar scaling relation is a formula for estimating some unknown property of a star from observations by scaling from the known properties of the Sun. 
These relations have the form 
\begin{equation} \label{eq:scaling}
    \frac{Y}{\text{Y}_\odot} 
    \simeq 
    \prod_i \left(\frac{X_i}{\text{X}_{\odot,i}}\right)^{P_i}
\end{equation}
where $Y$ is some property we wish to estimate, such as the radius of the star. 
The quantity $\text{Y}_\odot$ is the corresponding property of the Sun (e.g., the solar radius), the vector $\mathbf X$ contains measurable properties of the star (e.g., its effective temperature and luminosity), the vector $\mathbf X_{\odot}$ contains the corresponding solar properties, and $\mathbf P$ is some vector of exponents. 
An example scaling relation for estimating stellar radii can be derived from the Stefan--Boltzmann law as: 
\begin{equation}
    \frac{R}{\text{R}_\odot}
    \simeq 
    \left(
        \frac{T_{\text{eff}}}{\text{T}_{\text{eff},\odot}}
    \right)^{-2}
    \left(
        \frac{L}{\text{L}_\odot}
    \right)^{\frac{1}{2}}
\end{equation}
where $R$ is the radius, $T_{\text{eff}}$ the effective temperature, and $L$ the luminosity. 

In the era of space asteroseismology, relations known as the \emph{seismic scaling relations} have enjoyed wide usage. 
They are used to estimate the unknown masses and radii of stars by scaling asteroseismic observations with their helioseismic counterparts. 
These observations include the average frequency spacing between radial mode oscillations of consecutive radial order---the \emph{large frequency separation}, $\Delta\nu$---and the \emph{frequency at maximum oscillation power}, $\nu_{\max}$. 
From these and the observed effective temperature, one can estimate the stellar mass $M$ and radius $R$ of a star via: 
\begin{align}
    \frac{M}{\text{M}_\odot}
    &\simeq
    \bigg(
        \frac{\nu_{\max}}{\nu_{\max,\odot}}
    \bigg)^3
    \bigg(
        \frac{\Delta\nu}{\Delta\nu_\odot}
    \bigg)^{-4}
    \bigg(
        \frac{T_{\text{eff}}}{\text{T}_{\text{eff},\odot}}
    \bigg)^\frac{3}{2} \label{eq:scalingM}
\end{align}
\begin{align}
    \frac{R}{\text{R}_\odot}
    &\simeq
    \bigg(
        \frac{\nu_{\max}}{\nu_{\max,\odot}}
    \bigg)
    \bigg(
        \frac{\Delta\nu}{\Delta\nu_\odot}
    \bigg)^{-2}
    \bigg(
        \frac{T_{\text{eff}}}{\text{T}_{\text{eff},\odot}}
    \bigg)^\frac{1}{2} \label{eq:scalingR} 
\end{align}
where ${\nu_{\max,\odot} = 3090 \pm 30~\mu\text{Hz}}$, ${\Delta\nu_\odot = 135.1 \pm 0.1~\mu\text{Hz}}$ \mb{\citep{2011ApJ...743..143H}}, and ${\text{T}_{\text{eff},\odot} = 5772.0 \pm 0.8~\text{K}}$ \citep{2016AJ....152...41P}. 
These relations are useful thanks to the exquisite precision with which asteroseismic data can be obtained. 
For a typical well-observed solar-like star, $\Delta\nu$ and $\nu_{\max}$ can be measured with an estimated relative error of only 0.1\% and 1\%, respectively \citep[see, e.g., Figure~5 of][]{Bellinger2019}. 

These relations 
can be analytically derived from the fact that $\Delta\nu$ scales with the mean density of the star and $\nu_{\max}$ scales with the acoustic cut-off frequency \mb{\citep{1986ApJ...306L..37U, 1991ApJ...368..599B, 1995A&A...293...87K, Stello_2008, 2010A&A...509A..77K}}. 
\mbb{These relations are not perfectly accurate, however \citep[e.g.,][]{2009MNRAS.400L..80S, 2011A&A...530A.142B, 2016ApJ...832..121G, 2018MNRAS.478.4669T, 2019ApJ...870...41O}, especially when it comes to evolved stars, which has resulted in several suggested corrections \citep{2011ApJ...743..161W, 2016ApJ...822...15S, 2016MNRAS.460.4277G, 2017MNRAS.470.2069G}}. 

In recent years, there has been a great push for improved determination of stellar properties---and particularly stellar ages, for which asteroseismology is uniquely capable. 
Besides their intrinsic interest, knowledge of the ages of stars is useful for a broad spectrum of activities in astrophysics, ranging from charting the history of the Galaxy \mb{\citep[e.g.,][]{2013MNRAS.429..423M, doi:10.1146/annurev-astro-081915-023441, 2016MNRAS.455..987C, 2018MNRAS.475.5487S, 2018arXiv180900914S}} to understanding the processes of stellar and exoplanetary formation and evolution \mb{\citep[e.g.,][]{1981A&A....93..136M, 1996Natur.380..606L, doi:10.1146/annurev.earth.30.091201.140357, doi:10.1146/annurev-astro-081309-130806, 2015ARA&A..53..409W, 2017A&A...608A.112N, 2017ApJ...851...80B}}. 
Unlike with mass and radius, there is no scaling relation for stellar age. 
Instead, a multitude of methods have been developed for matching asteroseismic observations of stars to theoretical models of stellar evolution \mb{\citep[e.g.,][]{1994ApJ...427.1013B, 2004MNRAS.351..487P, 2009ApJ...700.1589S, Gai_2011, 2012MNRAS.427.1847B, 2014ApJS..214...27M, 2014A&A...569A..21L, 2014A&A...566A..82L, 2015MNRAS.452.2127S, 2017ApJ...835..173S, 2016ApJ...830...31B, 2017ApJ...839..116A, 2019MNRAS.484..771R}}, which then yields their ages. 

Scaling relations are attractive resources because they can easily and immediately be applied to observations without requiring access to theoretical models. 
In this paper, I seek to develop such a relation to estimate the ages of \mb{main sequence} stars, as well as to improve the scaling relations for estimating their mass and radius. 

The strategy is as follows. 
It is by now well-known that the core-hydrogen abundance (and, by proxy, the age) of a main-sequence star is correlated with the average frequency spacing between radial and quadrupole oscillation modes \citep[e.g.,][]{1984srps.conf...11C, 2010aste.book.....A, basuchaplin}. 
This spacing is known as the small frequency separation and is denoted by $\delta\nu$. 
The diagnostic power of $\delta\nu$ is owed to its sensitivity to the sound-speed gradient of the stellar core, which in turn is affected by the mean molecular weight, which increases monotonically over the main-sequence lifetime as a byproduct of hydrogen--helium fusion. 

Here I formulate new scaling relations that make use of this spacing $\delta\nu$, and I also add a term for metallicity. 
Rather than by analytic derivation, I calibrate the exponents of this relation using 80 solar-type stars whose ages and other parameters have been previously determined through detailed fits to stellar evolution simulations. 
Finally, I perform cross-validation to estimate the accuracy of the new relations. 

\section{Data}
\begin{figure}%
    \centering%
    \includegraphics[width=\linewidth]{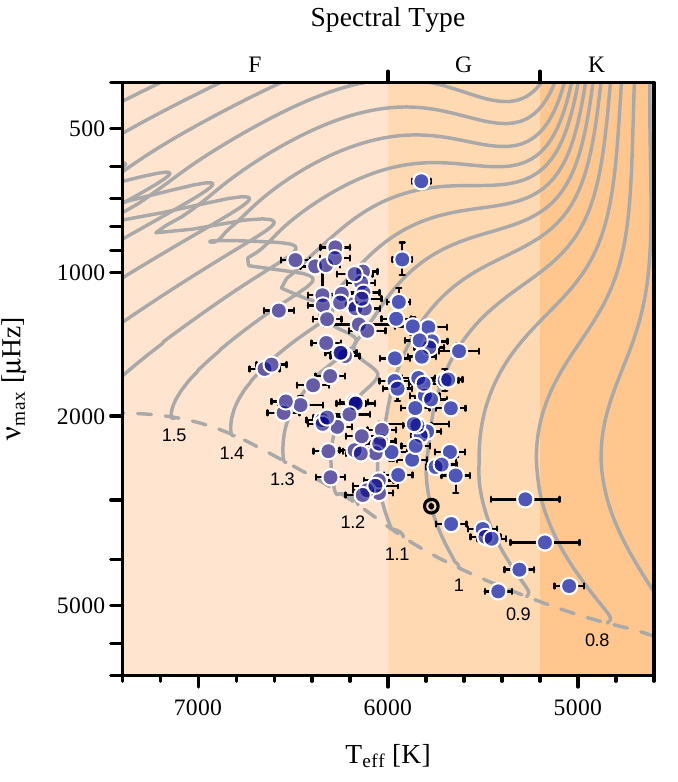}%
    \caption{\mb{Locations of the stars used in this study in the $\nu_{\max}$--$T_{\text{eff}}$ diagram. 
    The lines in the background are theoretical stellar evolution tracks of the indicated masses computed with \textsc{MESA} \citep{2011ApJS..192....3P,2013ApJS..208....4P,2015ApJS..220...15P,2018ApJS..234...34P}. 
    The zero-age main sequence is indicated with a dashed line; stars evolve upward and to the right. The background color indicates spectral type. The position of the Sun is indicated with the solar symbol ($\odot$).} \label{fig:teff-numax}}%
\end{figure}

I obtained spectroscopic and asteroseismic measurements of solar-like stars from the \emph{Kepler} Ages \citep{2015MNRAS.452.2127S, 2016MNRAS.456.2183D} and LEGACY \citep{2017ApJ...835..172L, 2017ApJ...835..173S} samples.
These stars were observed by the \emph{Kepler} spacecraft during its nominal mission 
\citep{2010Sci...327..977B}. 
\mb{These stars are all main-sequence or early sub-giant stars with no indications of mixed modes. Their positions in the $\nu_{\max}$--$T_{\text{eff}}$ plane are shown in Figure~\ref{fig:teff-numax}. 
The large and small frequency separations of these stars range from 38 to 180~$\mu$Hz and from 2.8 to 13~$\mu$Hz, respectively.}

The ages, masses, and radii of these stars are taken from \citet{Bellinger2019} as derived using the \emph{Stellar Parameters in an Instant} \citep[SPI,][]{2016ApJ...830...31B} pipeline. 
The SPI method uses machine learning to rapidly compute stellar ages, masses, and radii of stars by connecting their observations to theoretical models. 
For the present study, I selected 80 of these stars having $\delta\nu$ measurements with uncertainties smaller than 10\% and $\nu_{\max}$ measurements with uncertainties smaller than 5\%.

\section{Methods}
Here I will detail the construction of the new scaling relations and the procedure for their calibration. 
The scaling relations shown in Equations~\ref{eq:scalingM} and \ref{eq:scalingR} can be written more generically as follows. 
For a given quantity $Y$ (and corresponding solar quantity $\text{Y}_\odot$), 
\begin{align}
    \frac{Y}{\text{Y}_\odot}
    &\simeq
    \bigg(
        \frac{\nu_{\max}}{\nu_{\max,\odot}}
    \bigg)^\alpha
    \bigg(
        \frac{\Delta\nu}{\Delta\nu_\odot}
    \bigg)^\beta
    \bigg(
        \frac{\delta\nu}{\delta\nu_\odot}
    \bigg)^\gamma
    \bigg(
        \frac{T_{\text{eff}}}{T_{\text{eff},\odot}}
    \bigg)^\delta
    \exp\bigg(
        \text{[Fe/H]}
    \bigg)^\epsilon \label{eq:scalingX}
\end{align}
for suitable choices of the powers $\mathbf P = [\alpha, \beta, \gamma, \delta, \epsilon]$. 
Note the metallicity is first exponentiated. 
The uncertainties on all solar quantities are propagated except in the case of metallicity, where there is no agreed upon uncertainty \citep[e.g.,][]{2014dapb.book..245B}. 
From an analysis of solar data \citep{2014MNRAS.439.2025D} one can find ${\delta\nu_\odot = 8.957 \pm 0.059~\mu\text{Hz}}$.
For the solar age I use ${\tau_\odot = 4.569 \pm 0.006~\text{Gyr}}$ \citep{2015A&A...580A.130B}. 
To give a concrete example, in order to recover the classical radius scaling relation (Equation~\ref{eq:scalingR}), i.e., ${Y=R}$, we have ${\mathbf{P}=[1,-2,0,1/2,0]}$. 

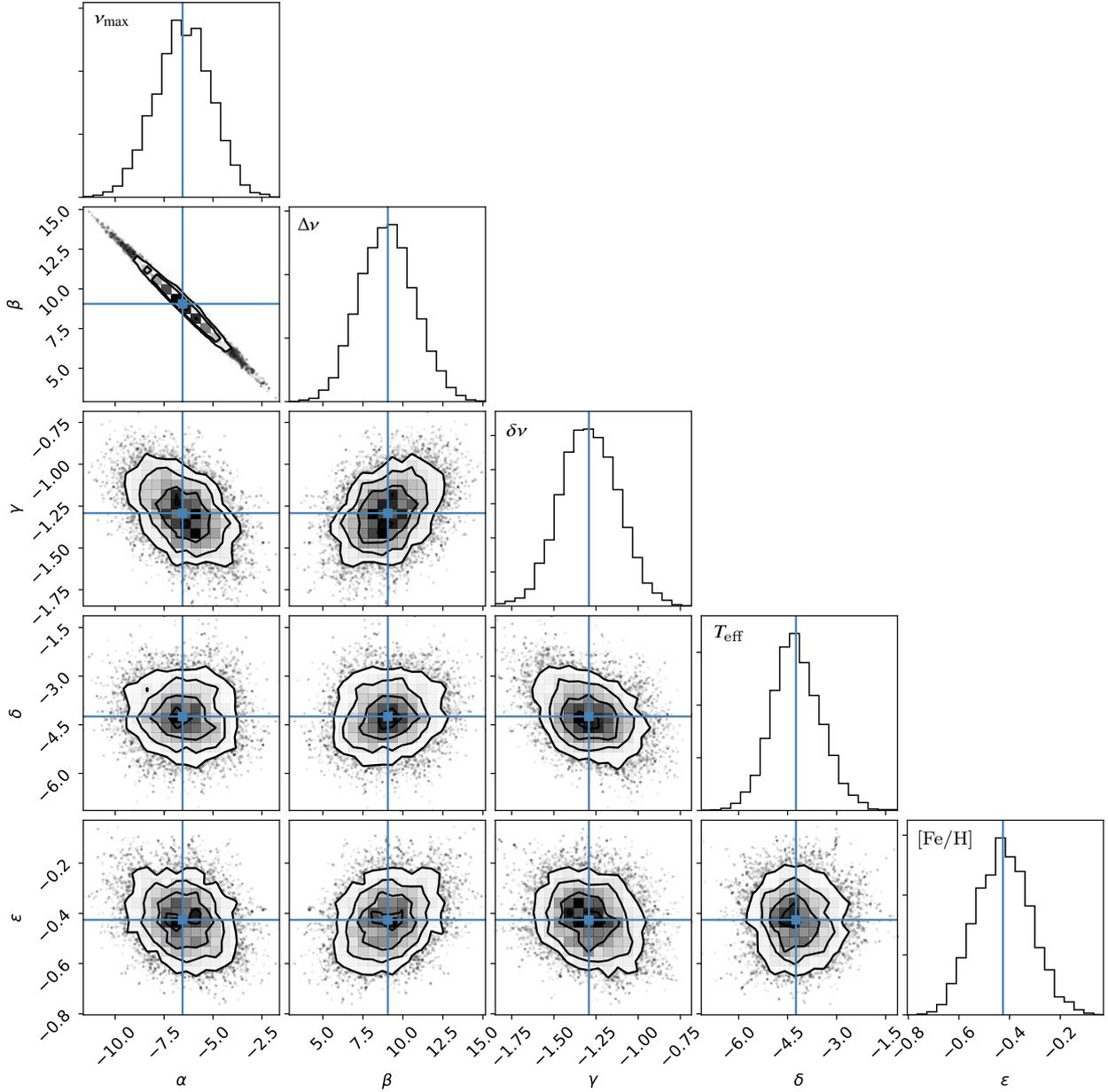
\begin{figure*}
    \centering
    \begin{overpic}[width=0.97\linewidth, tics=10, trim={0 0.4cm 0 1cm}, clip]{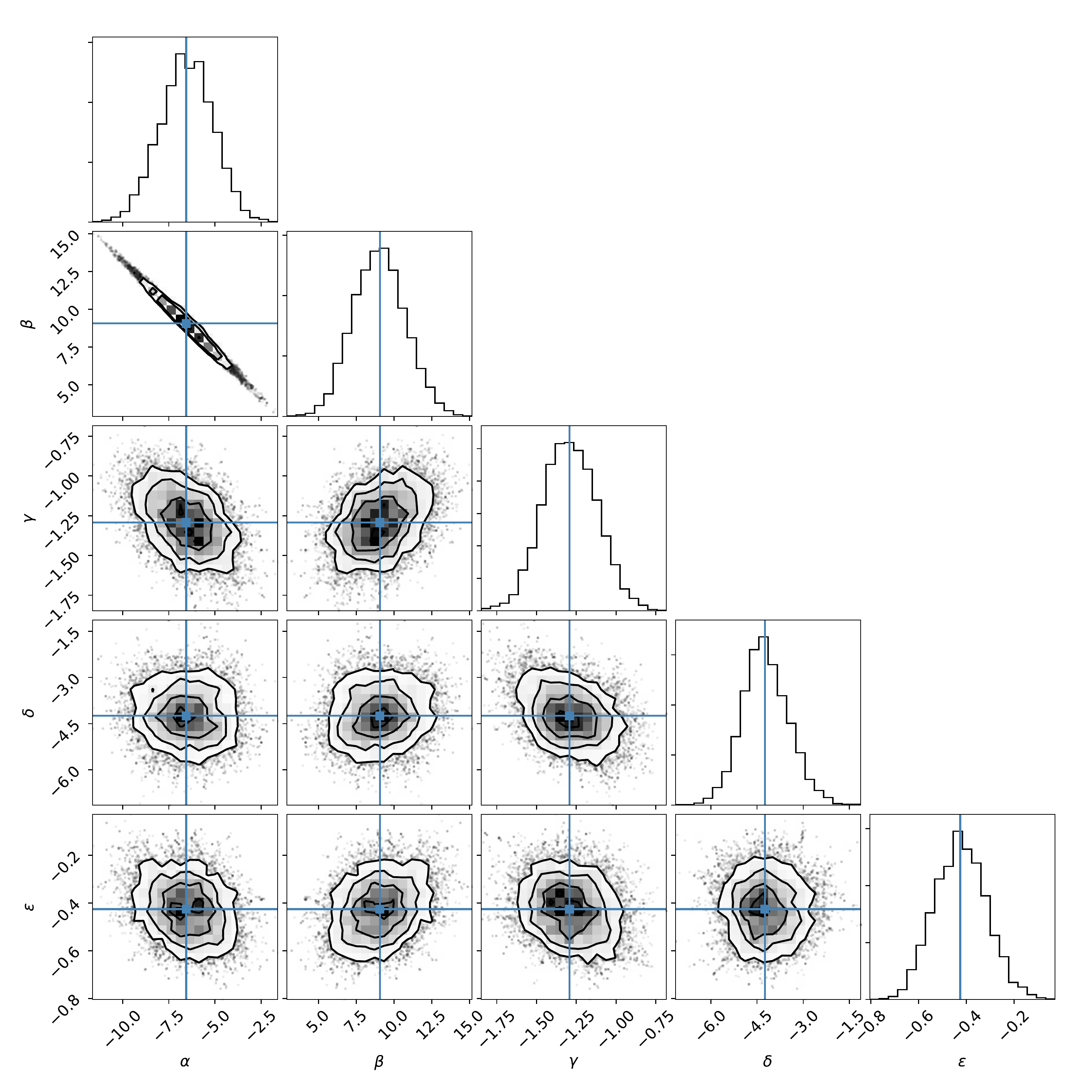}
        \put (9.5,93.5) {$\displaystyle\nu_{\max}$}
        \put (27,75.5) {$\displaystyle\Delta\nu$}
        \put (45,57.5) {$\displaystyle\delta\nu$}
        \put (63,40) {$\displaystyle{T}_{\text{eff}}$}
        \put (80.5,22) {$\displaystyle{\text{[Fe/H]}}$}
    \end{overpic}
    \caption{A standard corner plot showing the posterior distributions of each of the exponents in the MCMC-fitting of the age scaling relation (\emph{cf.}~Equation~\ref{eq:scalingX}). 
    Since $\nu_{\max}$ and $\Delta\nu$ are correlated quantities \citep[e.g.,][]{2009MNRAS.400L..80S}, the respective exponents for these quantities ($\alpha$ and $\beta$) are correlated as well. 
    The blue lines and points indicate the mode of the posterior distribution, which are listed in Table~\ref{tab:scaling}. 
    \label{fig:corner}}
\end{figure*}

I now seek to find the exponents for scaling relations that best match to the literature values of radius, mass, and age. 
I define the goodness-of-fit $\chi^2$ 
for a given vector $\mathbf P$ as
\begin{align}
    \chi^2 &= \sum_i \left(
        \frac{\hat Y_i - Y_i}{\sigma_i}
    \right)^2
\end{align}
where $\hat Y_i$ is the literature value of the desired quantity (e.g., age) for the $i$th star, $Y_i$ is the result of applying the scaling relation in Equation~\ref{eq:scalingX} with the given powers $\mathbf P$, and the uncertainties $\boldsymbol\sigma$ are the measurement uncertainty of $\boldsymbol{\hat{Y}}$. 
Given this function, I use Markov chain Monte Carlo \citep{2013PASP..125..306F, corner} to determine the posterior distribution of $\mathbf P$ for each of $Y=R, M, \tau$. 
Finally, I use the mean-shift algorithm \citep{fukunaga1975estimation, scikit-learn} to find the mode of the posterior distribution (see Figure~\ref{fig:corner}). 
This yields the desired exponents for each scaling law. 

\section{Results}
The best exponents for each of the computed scaling relations are shown in Table~\ref{tab:scaling}. 
I fit the mass and radius scaling relations both with and without a dependence on $\delta\nu$. I found that its inclusion made little difference, however, so I have omitted it for those two relations. 

\begin{table}
	\centering
	\caption{Classical and \mb{new} MCMC-fitted exponents for scaling relations (see Equation~\ref{eq:scalingX}). 
	\label{tab:scaling}}
	\begin{tabular}{ccccccc} 
		\hline
		          &  & $\nu_{\max}$ & $\Delta\nu$ & $\delta\nu$ & $T_{\text{eff}}$ & [Fe/H]\\\hline
		          & $Y$ & $\alpha$ & $\beta$ & $\gamma$ & $\delta$ & $\epsilon$\\\hline\hline
		Classic & $M$   &    3     &   -4    &   --    &    1.5     &    --  \\
		     New  & $M$   & 0.975 &   -1.435    &   --    &    1.216     &    0.270  \\\hline
		Classic & $R$  &    1     &   -2    &   --    &    0.5     &    --  \\
		     New  & $R$  & 0.305     &   -1.129 &   -- &    0.312 &    0.100 \\
         Seismic  & $R$  & 0.883     &   -1.859 & --      &   --     &   --     \\\hline
		     New  & Age  & -6.556 &   9.059    &   -1.292    &    -4.245     &    -0.426  \\
		\hline\hline
	\end{tabular}
\end{table}

A striking feature of the new radius scaling relation is its small dependence on spectroscopic variables. 
To explore this further, I have additionally fit a radius scaling relation using only $\Delta\nu$ and $\nu_{\max}$, \mb{referred to as `Seismic' in Table~\ref{tab:scaling}}. 
We shall soon see in the next section that although it is not quite as accurate as the full new radius relation, it does outperform the classical radius scaling relation, despite requiring no spectroscopic information. 

A notable aspect of the new mass and radius scaling relations is that their exponents are smaller in magnitude than those of the classical relations. 
As a consequence, the resulting uncertainties are also smaller. 
We may estimate the typical uncertainties of applying these relations by examining the typical uncertainties of their inputs. 
The median relative errors on $\nu_{\max}$, $\Delta\nu$, $\delta\nu$, $T_\text{eff}$, and $\exp\text{[Fe/H]}$ of the \emph{Kepler} Ages and LEGACY stars are approximately 1\%, 0.1\%, 4\%, 1\%, and 10\%, respectively \citep[e.g., Figure~5 of][]{Bellinger2019}. 
Thus, for a solar twin observed with such uncertainties, using Equation~\ref{eq:scalingX} with the exponents given in Table~\ref{tab:scaling} yields uncertainties of 0.032~$\text{M}_\odot$ (3.3\%), 0.011~$\text{R}_\odot$ (1.1\%), and 0.56~Gyr (12\%). 
These values are similar to those from fits to models \citep[e.g., Figure~6 of][]{Bellinger2019}.

\section{Benchmarking} \label{sec:cv}

I now seek to determine the accuracy of the relations: i.e., how well do these relations actually work? 
The fitted values cannot simply be compared to the literature values: it would be unsurprising if they match, being that they were numerically optimized to do so. 
Instead, we may use cross-validation to answer this question. 
The procedure is as follows. 
We take our same data set as before, but instead of training on the entire data set, we remove one of the stars. 
We then fit the relations using the other 79 stars that were not held out using the procedure described in the previous section. 
Finally, the newly fitted relations are tested on the held-out star. 
This test is then repeated for every star. 

\begin{figure*}
\vspace*{1.5\baselineskip}
    \centering%
    \begin{minipage}[t]{0.477\linewidth}%
    \includegraphics[width=\linewidth]{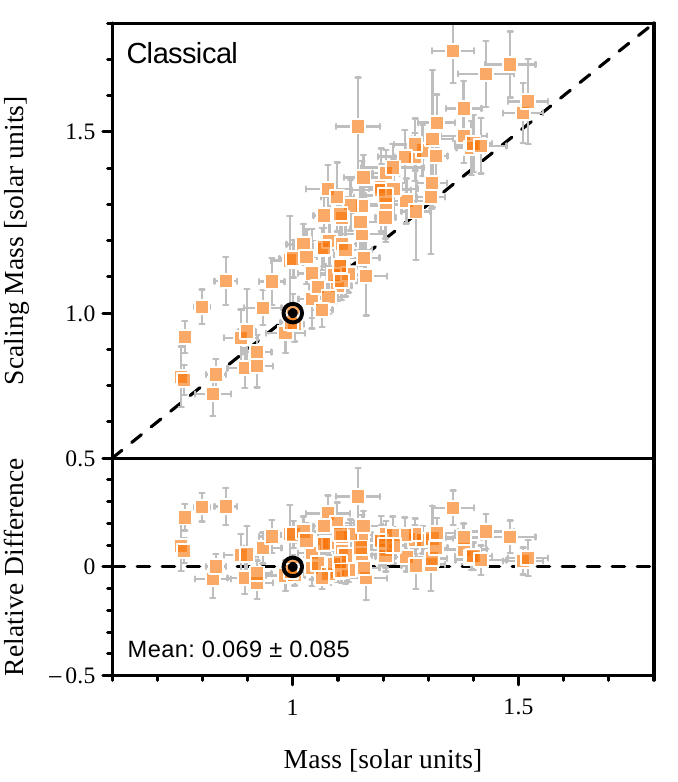}%
    \end{minipage}\hfill%
    \begin{minipage}[t]{0.477\linewidth}%
    \includegraphics[width=\linewidth]{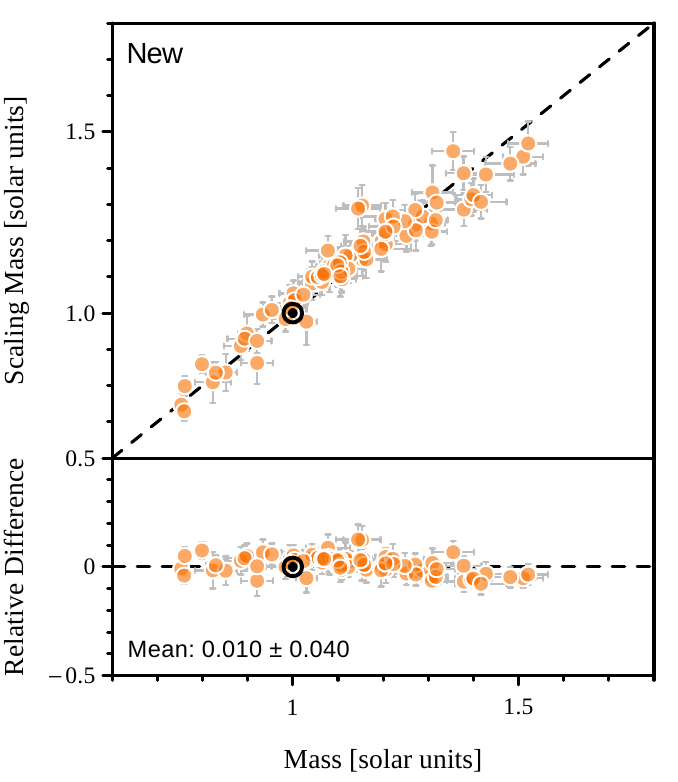}%
    \end{minipage}%
    \caption{Classical (left, squares) and new (right, circles) scaling relations for estimating stellar mass. 
    Each point is a star observed by \emph{Kepler}. 
    The masses on the x-axis \mb{are} the literature values; they were estimated using the SPI method with reference to a grid of theoretical stellar models. 
    The masses on the y-axis have been estimated using the scaling relation (whose fitting did not involve the star being plotted, see Section~\ref{sec:cv} for details). 
    The bottom panel shows the relative difference between the scaling mass and the literature mass. 
    The weighted mean and standard deviation of the ratios are given. 
    The Sun is denoted by the solar symbol ($\odot$). \label{fig:scalingM}}
\vspace*{1.5\baselineskip}
    \centering
    \begin{minipage}[t]{0.477\linewidth}%
    \includegraphics[width=\linewidth]{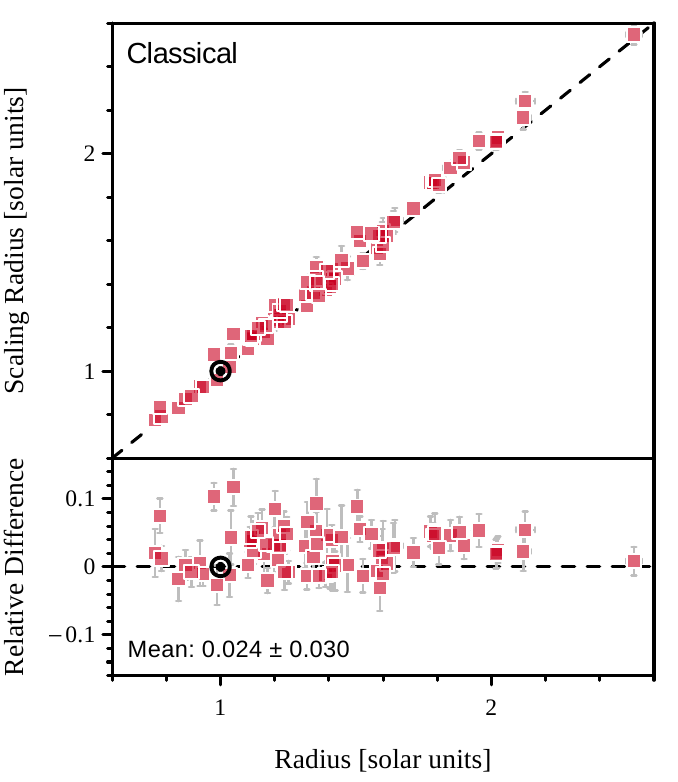}%
    \end{minipage}\hfill%
    \begin{minipage}[t]{0.477\linewidth}%
    \includegraphics[width=\linewidth]{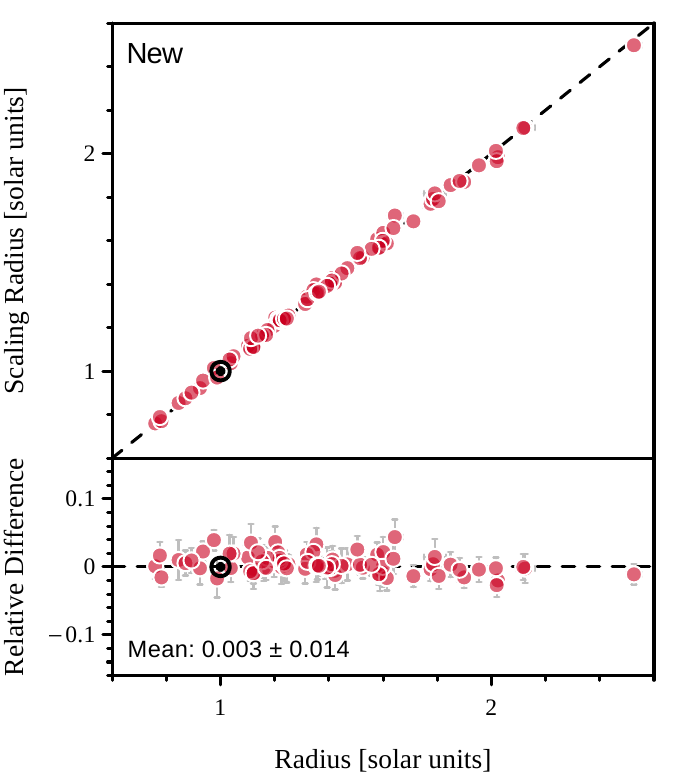}%
    \end{minipage}%
    \caption{Classical (left, squares) and new (right, circles) scaling relations for estimating stellar radius. \label{fig:scalingR}}
\end{figure*}

\begin{figure*}%
\vspace*{1.5\baselineskip}
    \centering%
    \includegraphics[width=0.477\linewidth]{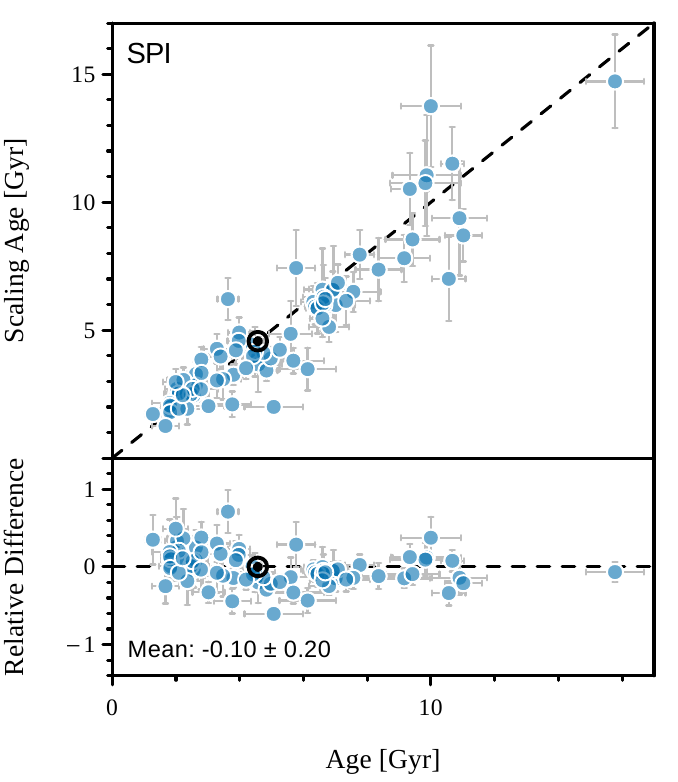}\hfill%
    \includegraphics[width=0.477\linewidth]{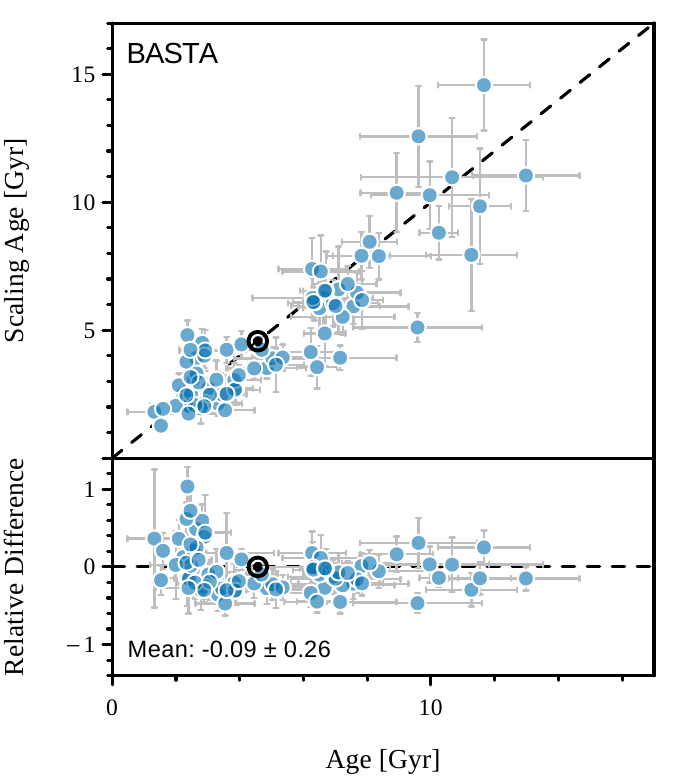}%
    \caption{A comparison of the scaling relation for stellar age to literature age values. Left panel: cross-validated scaling ages compared to literature values from SPI. Right panel: scaling ages using the values from Table~\ref{tab:scaling} in comparison with BASTA ages. 
    \label{fig:scaling-age}}%
\end{figure*}

\begin{figure}%
    \centering%
    \includegraphics[width=\linewidth]{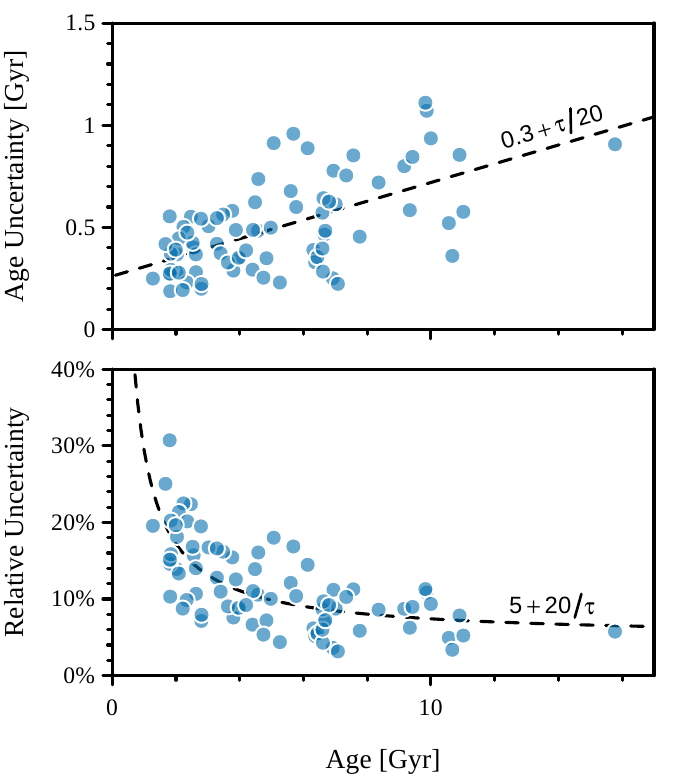}%
    \caption{The uncertainties of stellar ages from the literature as a function of age ($\tau$). Top panel: absolute uncertainties; bottom panel: relative uncertainties, in the sense of $\sigma_\tau / \tau$. Trend lines are shown to guide the eye. Older stars have more uncertain ages, in an absolute but not relative sense. \label{fig:unc-age}}%
\end{figure}

Comparisons of these cross-validated relations to literature values are shown in Figures~\ref{fig:scalingM}, \ref{fig:scalingR}, and \ref{fig:scaling-age}. 
The classical mass and radius scaling relations are also shown, and there it can be seen that the new relations are better at reproducing the literature values. 
The age scaling relation is also compared to the BASTA ages \citep{2015MNRAS.452.2127S, 2017ApJ...835..173S}, which were fit in a different way and using a different grid of theoretical models, \mb{and have some systematic differences with the SPI ages}. 
The age scaling relation shows a larger dispersion at older ages, which is most likely due to the input data having larger uncertainties there (see Figure~\ref{fig:unc-age}). 

\begin{figure*}%
\vspace*{1.5\baselineskip}
    \centering
    \begin{minipage}[t]{0.477\linewidth}
    \centering%
    \includegraphics[width=\linewidth]{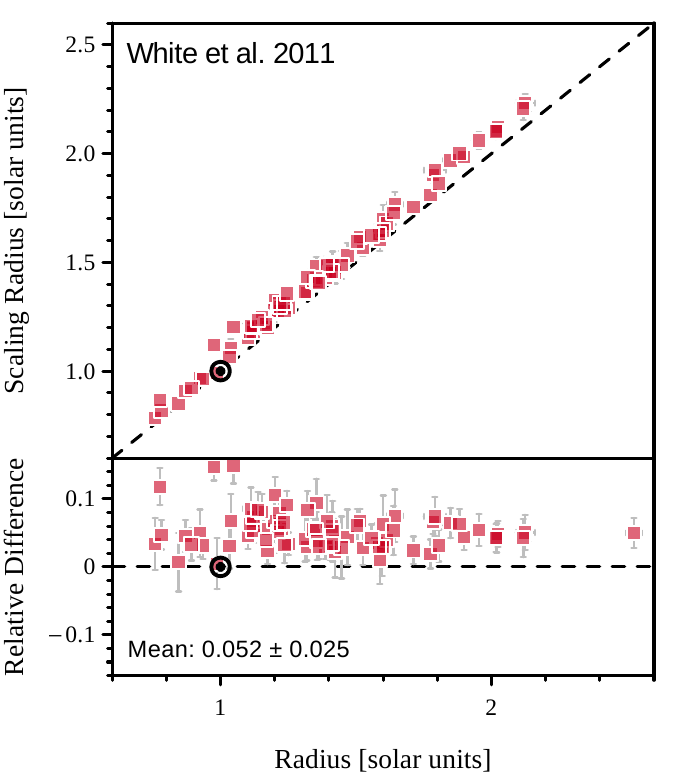}%
    \caption{Application of the \citet{2011ApJ...743..161W} radius scaling relation to the 80 stars. \label{fig:white}}%
    \end{minipage}\hfill%
    \begin{minipage}[t]{0.477\linewidth}
    \centering%
    \includegraphics[width=\linewidth]{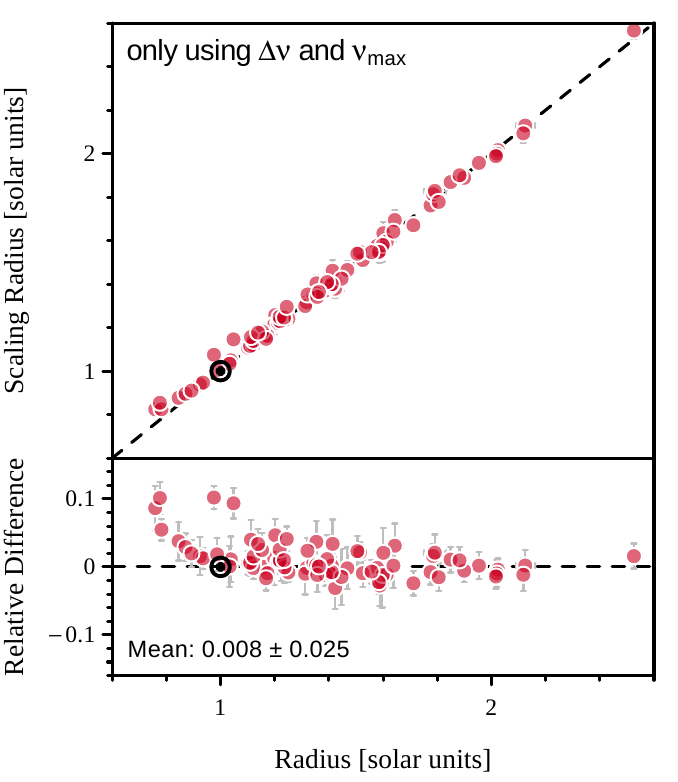}%
    \caption{Cross-validation of the purely seismic scaling relation for stellar radius (compare with the classical radius scaling relation in Figure~\ref{fig:scalingR}). \label{fig:scalingR-seismic}}%
    \end{minipage}\hfill
\end{figure*}

\mb{\citet{2011ApJ...743..161W} sought to improve the $\Delta\nu$ scaling relation by developing an analytical correction function from models. 
However, \citet{2018MNRAS.481L.125S} found that this correction actually degrades the agreement for main-sequence stars due to surface effects in the models. 
For comparison purposes, the \citet{2011ApJ...743..161W} radius scaling relation applied to these stars is shown in Figure~\ref{fig:white}.} 

The purely seismic radius scaling relation is shown in Figure~\ref{fig:scalingR-seismic}. 
\mb{There is a systematic trend at low radius; this is likely due to not including all relevant physics. 
Still, it similarly outperforms the classical scaling relation, despite having no temperature dependence. 
It is thus a potentially useful tool for measuring stellar radii without spectroscopy.} 


\begin{figure*}
\vspace*{1.5\baselineskip}
    \centering
    \includegraphics[width=0.477\linewidth]{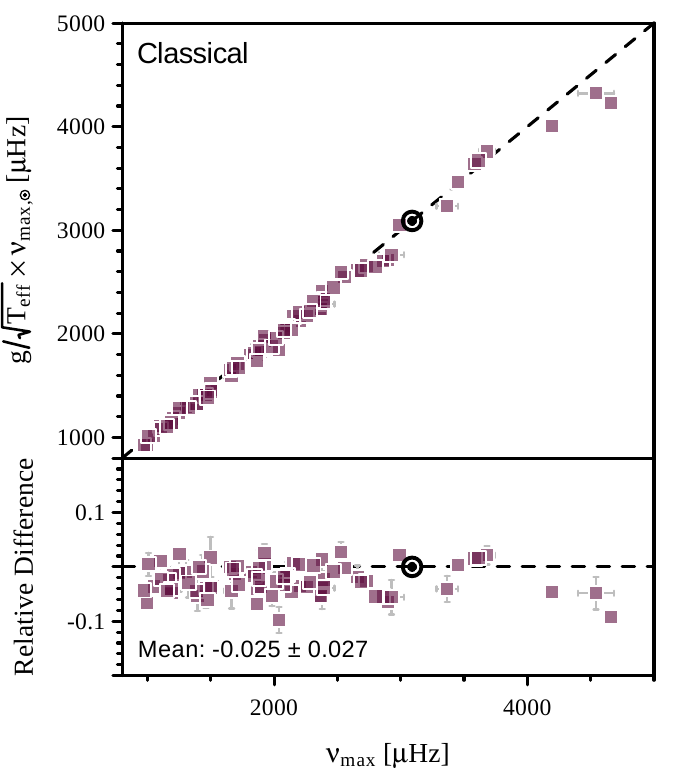}\hfill%
    \includegraphics[width=0.477\linewidth]{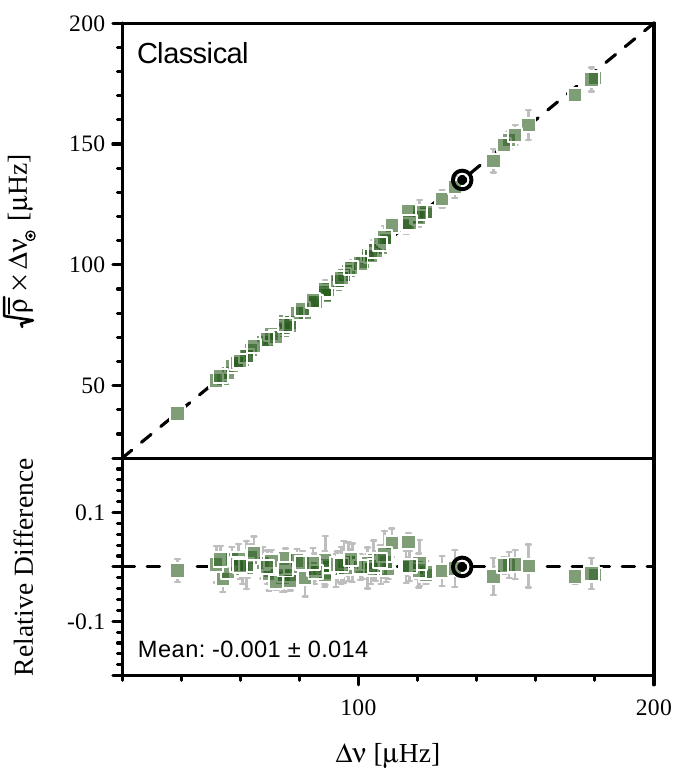}
    \caption{The $\nu_{\max}$ (left, Equation~\ref{eq:nu_max}) and $\Delta\nu$ (right, Equation~\ref{eq:Delta_nu}) scaling relations. %
    All quantities have been computed in solar units. %
    The surface gravities and mean densities are derived from the literature values of mass and radius; all other quantities are observed. %
    \label{fig:scaling}}
\end{figure*}

It is interesting to try to pinpoint the underlying cause of the discrepancies in the classical scaling relations seen in Figures~\ref{fig:scalingM} and \ref{fig:scalingR}. 
As mentioned, these relations come from
\begin{align}
    \nu_{\max} &\propto \nu_{\text{ac}} \propto g / \sqrt{T_{\text{eff}}} \label{eq:nu_max} \\
    \Delta\nu &\propto \sqrt{\bar\rho} \label{eq:Delta_nu}
\end{align}
where $\nu_{\text{ac}}$ is the acoustic cut-off frequency of the star, $g$ is its surface gravity and $\bar\rho$ is its mean density. 
Figure~\ref{fig:scaling} compares the left and right sides for both of these relations. 
While the $\Delta\nu$ scaling relation holds well (generally within about 1\%), the $\nu_{\max}$ scaling relation has larger scatter. 
This coincides with other recent findings that the classical $\nu_{\max}$ scaling relation should have additional dependencies \citep{2015A&A...583A..74J, 2017ApJ...843...11V}. 
A more accurate $\nu_{\max}$ relation can easily be inferred from the values given in Table~\ref{tab:scaling}. 

\begin{figure}%
    \centering%
    \includegraphics[width=\linewidth]{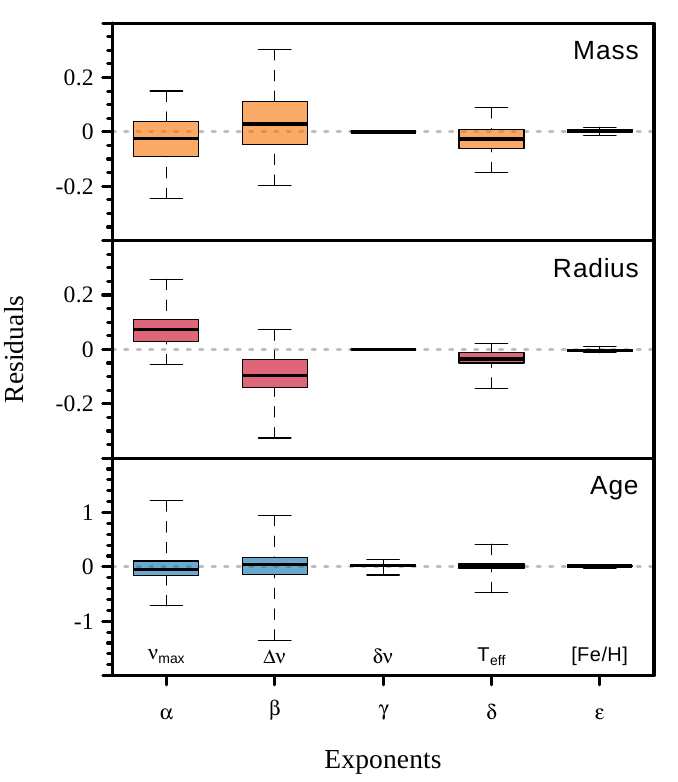}%
    \caption{Boxplots comparing the estimated exponents across the 80 cross-validation fits to the values given in Table~\ref{tab:scaling}, with zero being no difference. 
    The middle line shows the median of the residuals, the box shows the interquartile range, and the whiskers extend to the farthest points. 
    Note the differences in scale. 
    \label{fig:stability}}%
\end{figure}

\mb{The cross-validation} procedure may also be used to estimate the stability of the derived exponents. 
Figure~\ref{fig:stability} compares the exponents for each of the 80 fits when holding out one star each time. 
The exponents do not change much between the different fits, which indicates convergence. 

\begin{figure*}
\vspace*{1.5\baselineskip}
    \centering%
    \begin{minipage}[t]{0.477\linewidth}%
    \includegraphics[width=\linewidth]{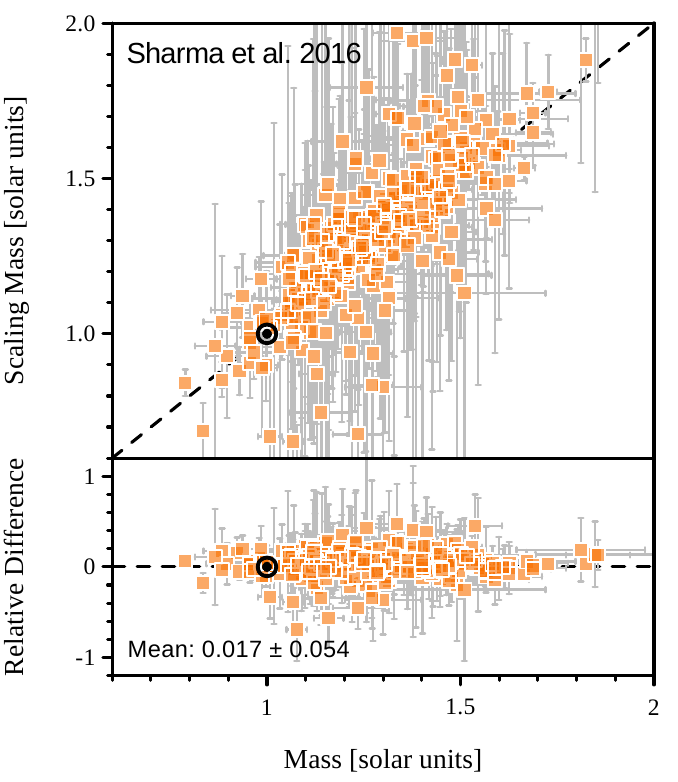}%
    \end{minipage}\hfill%
    \begin{minipage}[t]{0.477\linewidth}%
    \includegraphics[width=\linewidth]{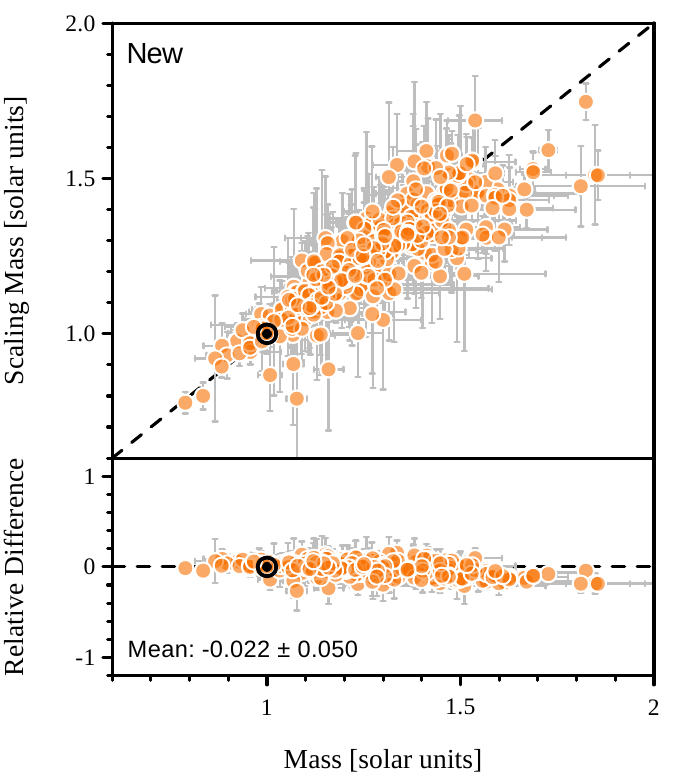}%
    \end{minipage}%
    \caption{\mbb{Comparison of \citealt{2016ApJ...822...15S} (left, squares) and new (right, circles) scaling relations as applied to the APOKASC sample of 408 main-sequence and sub-giant stars. The x-axis shows the masses of these stars as given in the APOKASC catalogue, which were determined through grid-based modelling.} \label{fig:APOKASC-M}}
\vspace*{1.5\baselineskip}
    \centering
    \begin{minipage}[t]{0.477\linewidth}%
    \includegraphics[width=\linewidth]{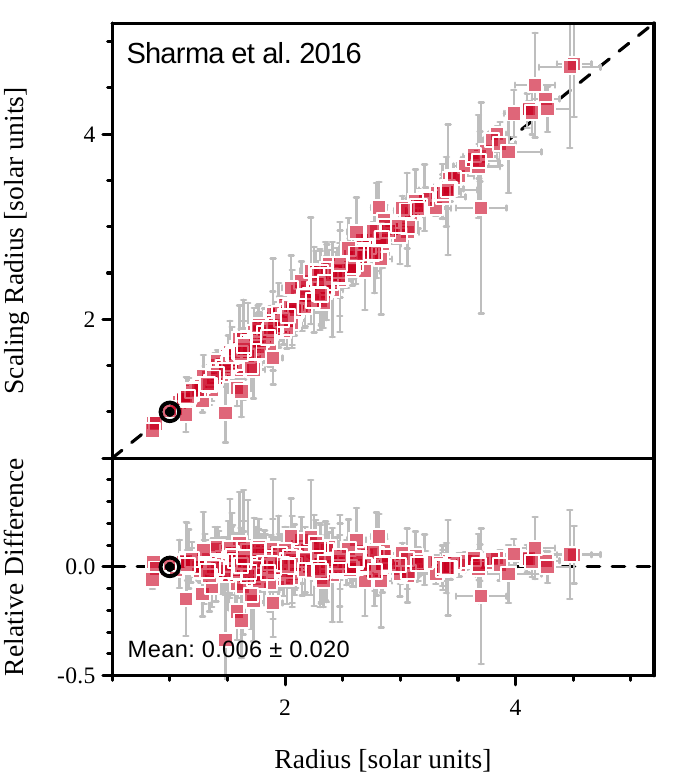}%
    \end{minipage}\hfill%
    \begin{minipage}[t]{0.477\linewidth}%
    \includegraphics[width=\linewidth]{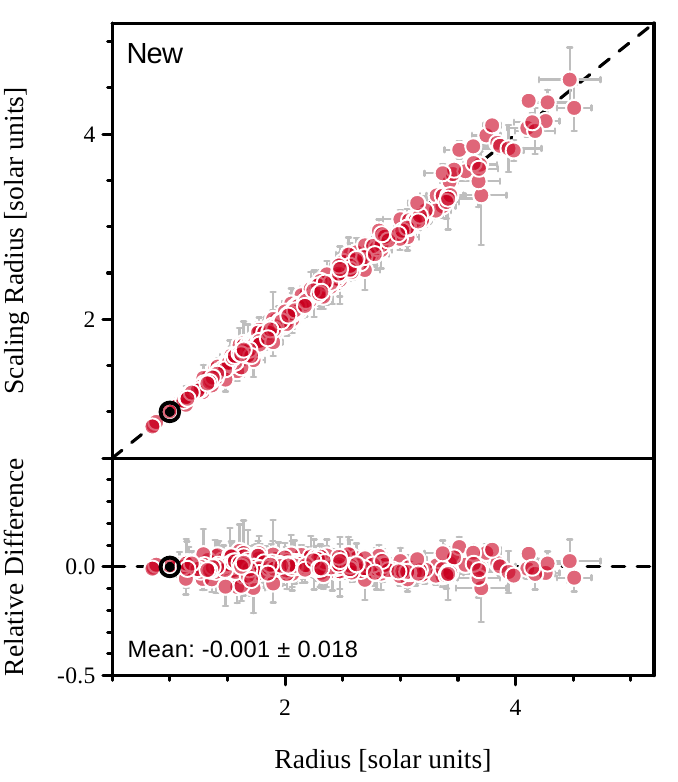}%
    \end{minipage}%
    \caption{\mbb{\citealt{2016ApJ...822...15S} (left, squares) and new (right, circles) scaling relations for estimating stellar radius applied to the same APOKASC sample of stars as in Figure~\ref{fig:APOKASC-M}.} \label{fig:APOKASC-R}}
\end{figure*}

\mb{As a final test, I have applied the new mass and radius scaling relations to the APOKASC+SDSS sample of dwarfs and sub-giant stars \citep{2017ApJS..233...23S}. 
This sample includes more evolved stars than those in the training set, with $\nu_{\max}$ and $\Delta\nu$ values down to 250 and 17 $\mu$Hz, respectively. 
\mbb{Figures~\ref{fig:APOKASC-M} and \ref{fig:APOKASC-R} show that} the new radius, purely seismic radius, and new mass scaling solutions for these stars match the grid-based modeling results on average within 1.8\%, 3.7\%, and 5\%, respectively. 
Being that the average uncertainties of these stars from modelling are 2.4\% in radius and 4.2\% in mass, the new relations give solutions that are of order or beneath typical uncertainties in modelling. 
\mbb{Comparisons with the \citet{2016ApJ...822...15S} scaling relations, which are calculated by interpolating in a grid of models, are shown in those figures as well. 
It can be seen that the new scaling relations have less scatter and smaller uncertainties.}} 

\section{Discussion \& Conclusions} 
In this paper, I formulated new scaling relations for estimating the mass, radius, and age of solar-like stars. 
I calibrated the free parameters of these relations using 80 well-studied stars whose ages have been previously determined from fits to theoretical stellar models. 
The values for the calibrated relations are listed in Table~\ref{tab:scaling}. 
I used cross-validation to gauge the stability of the exponents of the relations, and also to assess their accuracy. 
The relations were found to have a typical precision that is similar to methods which make reference to a grid of theoretical models. 
Finally, when compared with the classical scaling relations and other proposed corrections, the new relations were found to be in better agreement with the literature values of mass and radius. 
For easy use, source code for these relations can be found in the \hyperref[sec:code]{Appendix}.

A few points of discussion are in order. 
These relations have been fit to literature values of age, mass, and radius, which themselves were determined via fits to a grid of theoretical stellar models. 
Therefore, these relations are model-dependent. 
It follows that errors in the literature values may affect the accuracy of these relations. 
Several sources of systematic errors may exist in the theoretical models used to estimate stellar ages, such as unpropagated uncertainties in nuclear reaction rates or atmospheric abundances. 
When stellar models are inevitably improved, the stars should be fit again, and these relations re-calibrated. 

That being said, the literature values used to calibrate these relations have been found to be in good agreement with external constraints, such as \emph{Gaia} radii and luminosities, interferometry, and also with other modeling efforts which are based on different theoretical models \citep{2016ApJ...830...31B, Bellinger2019}. 
\mb{A large effort was made in those works to propagate sources of uncertainty stemming from the unknown model physics inputs of diffusion, convective mixing length, and convective over- and under-shooting.} 

One might be tempted to calibrate these relations with stellar models directly, circumventing the need to use real stars whose ages have been fit to said models. 
However, it must be kept in mind that theoretical values of $\nu_{\max}$ are generally computed using the scaling relation. 
Furthermore, theoretical calculations of the large and small frequency separation are affected by the asteroseismic surface term, giving rise to systematic discrepancies between theory and observation. 
The ages used here were instead determined using asteroseismic frequency ratios, which are insensitive to surface effects \citep{2003A&A...411..215R, 2005MNRAS.356..671O}, but more difficult for observers to measure. 
This approach allows us the convenience of using the observed large and small frequency separations and $\nu_{\max}$. 

\mb{While this paper constitutes the first effort, as far as I am aware, to develop a scaling relation for stellar age, several previous studies have sought to improve the mass and radius scaling relations. 
Generally the focus has been on red-giant stars, where the discrepancies are most apparent. 
A few approaches have been tried. 
\citet{2011ApJ...743..161W} and \citet{2016MNRAS.460.4277G, 2017MNRAS.470.2069G} developed analytic functions to correct for deviations in the $\Delta\nu$ scaling based on stellar models. 
In another approach, \citet{2016ApJ...822...15S} interpolated corrections based on a grid of models. 
\citet{2018A&A...616A.104K} developed empirical functions fitting to six red giants with orbital measurements of mass and radius.}

\mb{One distinction, apart from accuracy, is that the relations developed here are not arbitrary in their form; instead, they are explicit solar homology relations following Equation~\ref{eq:scaling}. 
The tests presented in this manuscript demonstrate that the relations work well within the tested ranges (i.e., on main-sequence and early sub-giant stars). 
Extrapolation of these relations (i.e., to late sub-giant and red-giant stars) is not recommended; the development of new scaling relations for red giants is to be explored in a future work, at which point it will be interesting to compare with the red-giant correction functions. 
It will also be interesting to calibrate a new age relation to such stars, as there the small frequency separation ceases to be a useful diagnostic.}

\section*{Acknowledgements}
The author thanks Hans Kjeldsen, J{\o}rgen Christensen-Dalsgaard, and the anonymous referee for their suggestions which have improved the manuscript. 
Funding for the Stellar Astrophysics Centre is provided by The Danish National Research Foundation (Grant agreement no.: DNRF106). 
The numerical results presented in this work were obtained at the Centre for Scientific Computing, Aarhus\footnote{\url{http://phys.au.dk/forskning/cscaa/}}.


\bibliographystyle{mnras}
\bibliography{Bellinger} 



\clearpage
\appendix\onecolumn
\section{Source code} \label{sec:code}
For the sake of convenience, here I provide source code in Python 3 to make use of these new scaling relations. \vspace*{0.7\baselineskip}

\begin{myminted}{scaling.py}
from math import e
from uncertainties import ufloat

# Enter some data, for example a solar twin 
# ufloat holds a measurement and its uncertainty 
nu_max   = ufloat(3090,  3090  * 0.01)  # muHz, with 1%   uncertainty 
Delta_nu = ufloat(135.1, 135.1 * 0.001) # muHz, with 0.1% uncertainty 
delta_nu = ufloat(8.957, 8.957 * 0.04)  # muHz, with 4%   uncertainty 
Teff     = ufloat(5772,  5772  * 0.01)  # K,    with 1%   uncertainty 
Fe_H     = ufloat(0,             0.1)   # dex,  0.1 dex   uncertainty 

# Take the powers from Table 1, here given with more precision 
# P        = [      alpha,        beta,        gamma,       delta,     epsilon]
P_age      = [-6.55598425,  9.05883854, -1.29229053, -4.24528340, -0.42594767]
P_mass     = [ 0.97531880, -1.43472745,  0,           1.21647950,  0.27014278]
P_radius   = [ 0.30490057, -1.12949647,  0,           0.31236570,  0.10024562]
P_R_seis   = [ 0.88364851, -1.85899352,  0,           0,           0         ]

# Apply the scaling relation 
def scaling(nu_max, Delta_nu, delta_nu, Teff, exp_Fe_H, P=P_age, 
            nu_max_Sun = ufloat(3090,  30),    # muHz
          Delta_nu_Sun = ufloat(135.1, 0.1),   # muHz 
          delta_nu_Sun = ufloat(8.957, 0.059), # muHz
              Teff_Sun = ufloat(5772,  0.8),   # K
              Fe_H_Sun = ufloat(0,     0)):    # dex
    
    alpha, beta, gamma, delta, epsilon = P 
    
    # Equation 5 
    return ((nu_max   /   nu_max_Sun) ** alpha *
            (Delta_nu / Delta_nu_Sun) ** beta  *
            (delta_nu / delta_nu_Sun) ** gamma * 
            (Teff     /     Teff_Sun) ** delta * 
            (e**Fe_H  /  e**Fe_H_Sun) ** epsilon)

scaling_mass   = scaling(nu_max, Delta_nu, delta_nu, Teff, Fe_H, P=P_mass)
scaling_radius = scaling(nu_max, Delta_nu, delta_nu, Teff, Fe_H, P=P_radius)
scaling_age    = scaling(nu_max, Delta_nu, delta_nu, Teff, Fe_H, P=P_age) * \
    ufloat(4.569, 0.006)

print('Mass:',   scaling_mass,   '[solar units]')
print('Radius:', scaling_radius, '[solar units]')
print('Age:', '{:.2u}'.format(scaling_age), '[Gyr]')
\end{myminted}
\begin{myminted2}
$ python3 scaling.py
Mass: 1.000+/-0.033 [solar units]
Radius: 1.000+/-0.011 [solar units]
Age: 4.57+/-0.56 [Gyr]
\end{myminted2}

\bsp	
\label{lastpage}
\end{document}